\documentstyle[amssymb,aps,12pt]{revtex}

\begin{document}
\author{\thinspace M. J. Steel and D. F. Walls}
\address{Department of Physics, University of Auckland, Private Bag 92019, Auckland,\\
New Zealand.}
\title{Pumping of twin-trap Bose-Einstein condensates}
\date{\today}
\maketitle

\begin{abstract}
We consider extensions of the twin-trap Bose-Einstein condensate system of
Javaneinen and Yoo [Phys. Rev. Lett., {\bf 76}, 161--164 (1996)] to include
pumping and output couplers. Such a system permits a continual outflow of
two beams of atoms with a relative phase coherence maintained by the
detection process. We study this system for two forms of thermal pumping,
both with and without the influence of inter-atomic collisions. We also
examine the effects of pumping on the phenomenon of collapses and revivals
of the relative phase between the condensates.
\end{abstract}

\newpage

\section{Introduction}

Both before and since the recent demonstrations of Bose-Einstein
condensation~\cite{and95,dav95,bra97} in dilute alkali gases, the concept of
the phase of a Bose-Einstein condensate (BEC) has attracted a deal of
theoretical study. Traditionally, the existence of a phase is taken as a
signature of spontaneous symmetry breaking and strictly it is only the
relative phase of two BEC's that can be assigned a definite value. Many
people have discussed the difficulties associated with the fact that in many
cases, we may know the number of atoms present fairly precisely. The
condensate is thus in a state that cannot possess a well-defined phase.
Recently, several authors~\cite{jav96,cir96,nar96,won96,won96a,jac96} have
modelled systems containing two BEC's, demonstrating that a relative phase
can arise naturally, even when both condensates are initially in number
states. Typically, atoms are permitted to leak out from both traps and are
detected by some apparatus below. As it is unknown from which trap any
individual atom comes, the distribution of positions at which atoms are
detected shows interference fringes. At the same time, the quantum state of
the two traps evolves from a simple product of number states into an
entangled state of varying number {\em differences} between the traps
allowing a well-defined relative phase to appear between the two
condensates. The value of the relative phase is randomly distributed from
run to run so that for an ensemble of runs, the phase symmetry is restored.

The influence of inter-atomic collisions on such a system has also been
considered~\cite{won96,won96a,wri96,wri97}. The notable results are a
reduction in the visibility of the observed interference pattern, and in the
conditional visibility of the entangled state, due to diffusion of the
condensate phase. Moreover, stopping the detection process after a relative
phase has been established leads to collapses and revivals in the
conditional visibility~\cite{won96a,wri96,wri97}, as the collisions cause
the phases of different components of the entangled state to rotate at
different rates. As the total state is a discrete sum of number difference
states, the phases realign periodically and the visibility is restored.

An obvious consequence of detecting atoms is that the system can not attain
a steady state. In time, the trap occupancies fall so low that the entangled
state is reduced in size until eventually the atoms in one or both traps are
exhausted altogether. If one envisaged using the two-trap system as a real
``device'' rather than merely an analogy for a single condensate, this might
be a serious problem. A natural use would be as a kind of two-beam ``atom
laser'' in which additional atoms would be tapped off from each trap through
an output coupler into separate beams\ (see Fig.~\ref{fig:geom}), as in the
case of two remarkable recent experiments at MIT~\cite{mew97,and97}. The
relative phase between the BEC's in the traps built up by the measurement
process would then be reflected in the same well-defined relative phase
between the two output beams, that could be exploited for some other
interference experiment. While such a scheme could operate in a pulsed
fashion where the traps were repeatedly filled with atoms, measured until
exhausted and then the process repeated (as was the case for the experiments
just mentioned), it would also be useful to have a continuous output. This
would clearly require a continuous pumping of new atoms into the traps. We
note of course, that interference measurements on the output beams would
themselves act to produce a relative phase between the traps. It may be
however, that the desired rate of measurements on the output beams is too
low to stabilize the phase over long periods, whereas the detection rate
directly on the traps may be as large as necessary. Here we assume that the
rate at which subsequent measurements are made on the output beams is so low
that they have a negligibly small feedback on the entangled state in the
traps. (Note that the output rate may be relatively high as long as most of
the leaked atoms are not actually detected in an interference experiment.)

In this paper, we explore the effects of pumping a two-trap BEC system. We
investigate what kind of steady states may be reached when pumping is
included, and study the competing effects of the collisions and measurements
in such a steady state. We also allow output couplers on each trap as
discussed above. We explore two types of pumping from thermal atom baths
coupled to each trap. In the first, we allow two-way pumping where atoms are
exchanged with the baths in both directions. Such a scheme was considered
for a single-trap atom laser based on evaporative cooling by Wiseman ${\em et%
}${\em \ al~}\cite{wis96}. In the second model, atoms may enter the traps
from a thermal bath, but the reverse process is forbidden. This kind of
irreversible pumping scheme has been considered in an atom laser model by
Holland {\em et al~}\cite{hol96a}. We can then consider our whole system as
a means of transferring non-condensate atoms in a thermal bath, into two
coherent beams with the coherence established by the detection process. In
this sense, the system might be considered a primitive two-beam atom laser.
We emphasize though, that in this paper we include no line-narrowing
element, a central component of true optical lasers~\cite{sar74,hak84,wis97}%
. The linewidth of the output beams would be at least that of the output
couplers.

The paper is structured as follows. In section~\ref{sec:ingreds} we describe
our model in detail while in section~\ref{sec:state} we discuss the nature
of the entangled state more fully. Using Monte-Carlo wave function
simulations, we consider pumping from a thermal bath of atoms in section~\ref
{sec:steady} examining the visibility and other parameters, when the pumping
of the trap is either two-way or inwards-only . Finally in section~\ref
{sec:collapses}, we turn to the phenomenon of collapses and revivals of the
condensate phase and present some interesting effects associated with
pumping.

\section{\protect\smallskip Elements of the model}

\label{sec:ingreds}

We now set out the model in full detail. In all the work below we assume a
system of two traps containing condensates with occupation numbers $n_{1}$
and $n_{2}$ (see Fig.~\ref{fig:geom}). We occasionally write $n_{i}$ to
indicate either trap. Atoms may leak from either trap and be detected below
at a mean rate $\gamma n_{i}$ establishing a relative phase. (We assume that
an atom in either trap has the same probability of detection). The traps are
pumped from two thermal reservoirs containing $N_{1}$ and $N_{2}$
non-condensate atoms with rate coefficients $\chi _{1}$ and $\chi _{2}$
respectively. At different points in the paper, we assume that atoms may
move either in both directions between the traps and reservoirs, or only
into the traps, so that we define separate rates $\chi _{1}^{\text{in}},$ $%
\chi _{2}^{\text{in}}$ and $\chi _{1}^{\text{out}}$ and $\chi _{2}^{\text{out%
}}$. In the simplest systems, we would expect $\chi _{i}^{\text{in}}=\chi
_{i}^{\text{out}},$ but the inclusion of some irreversible pumping process
can prevent atoms escaping from the trap into the baths giving $\chi _{i}^{%
\text{out}}=0.$ One example would be to couple the thermal bath to an
excited trap level $\left| e\right\rangle $. An atom in $\left|
e\right\rangle $ can decay to the BEC\ mode $\left| g\right\rangle $ by
spontaneous emission. If the medium is optically thin, the emitted photon is
lost and excitation out of the ground state is impossible. We see in section~%
\ref{sec:steady} that this one-way pumping leads to quite different behavior
to the two-way pumping. The physical validity of the two types of thermal
pumping has been discussed at some length in Ref.~\cite{wis96}.

We also allow a separate leak from each trap into an empty mode at rates $%
\nu _{1}$ and $\nu _{2}$ to act as ``output'' beams. While we here leave the
mechanism unspecified, we note that several techniques for creating an
output coupler have been demonstrated by the MIT group~\cite{mew97} in which
rf signals or a bias in the trapping field are used to couple a portion of
the condensate to untrapped spin states~\cite{bal97}. Finally, the atoms in
each trap experience collisions amongst themselves at a rate $\kappa .$

The master equation for the system may thus be written 
\begin{eqnarray}
\frac{d\rho }{dt} &=&\frac{i}{\hbar }\left[ \rho ,H\right] +\gamma
\int_{0}^{2\pi }D\left[ \Psi \left( \phi \right) \right] \,d\phi \,\rho
+\chi _{1}^{\text{out}}\left( N_{1}+1\right) D\left[ a_{1}\right] \rho +\chi
_{1}^{\text{in}}N_{1}D\left[ a_{1}^{\dag }\right] \rho  \nonumber \\
&&+\chi _{2}^{\text{out}}\left( N_{2}+1\right) D\left[ a_{2}\right] \rho
+\chi _{2}^{\text{in}}N_{2}D\left[ a_{2}^{\dag }\right] \rho +\nu
_{1}D\left[ a_{1}\right] \rho +\nu _{2}D\left[ a_{2}\right] \rho ,
\label{eq:master}
\end{eqnarray}
where the Hamiltonian describing collisions amongst the atoms is~\cite{wri97}
\begin{equation}
H=\frac{\kappa }{2}\left[ \left( a_{1}^{\dag }a_{1}\right) ^{2}+\left(
a_{2}^{\dag }a_{2}\right) ^{2}\right] ,  \label{eq:hamil}
\end{equation}
and $a_{i}$ is the annihilation operator for an atom in trap $i.$ Also, for
an arbitrary operator $c,$ the superoperator $D\left[ c\right] \rho $ is
defined by 
\begin{equation}
D\left[ c\right] \rho =c\rho c^{\dag }-\frac{1}{2}\left( c^{\dag }c\rho
+\rho c^{\dag }c\right) .
\end{equation}
The field operator $\psi \left( \phi \right) =a_{1}+a_{2}e^{-i\phi },$ where 
$\phi =2\pi x,$ describes the detection of an atom at position $x.$ Most of
our results are obtained from Monte-Carlo wave-function simulations of Eq.~(%
\ref{eq:master}) in which all leaks and additions of atoms to the traps are
represented as quantum jumps, and the non-unitary evolution of the wave
function is given by the effective Hamiltonian 
\begin{eqnarray}
H_{\text{eff }} &=&H-i\frac{\hbar }{2}\left[ \gamma \left( a_{1}^{\dag
}a_{1}+a_{2}^{\dag }a_{2}\right) +\left[ \chi _{1}^{\text{out}}\left(
N_{1}+1\right) +\chi _{1}^{\text{in}}N_{1}+\nu _{1}\right] a_{1}^{\dag
}a_{1}\right. \\
&&\left. +\chi _{1}^{\text{in}}N_{1}+\left[ \chi _{2}^{\text{out}}\left(
N_{2}+1\right) +\chi _{2}^{\text{in}}N_{2}+\nu _{2}\right] \ a_{2}^{\dag
}a_{2}+\chi _{2}^{\text{in}}N_{2}\right] .
\end{eqnarray}
When the chosen jump is a detection, the phase $\phi $ of the next detection
is chosen randomly according to the conditional probability distribution 
\begin{equation}
P\left( \phi \right) =\left\langle t\right| \psi ^{\dag }\left( \phi \right)
\psi \left( \phi \right) \left| t\right\rangle ,
\end{equation}
where $\left| t\right\rangle $ is the state of the system immediately
preceding the detection. It has been shown elsewhere~\cite
{jav96,won96,won96a} that this may always be written in the form

\begin{equation}
P\left( \phi \right) \propto 1+\beta \cos \left( \phi +\theta \right) ,
\label{eq:condprob}
\end{equation}
where the conditional visibility $\beta $ and conditional phase $\theta $
are determined by the previous history of the system.

In order that the average number of atoms in the traps become constant in
time for the thermal pumping schemes, we require a relation between the
various coefficients in the master equation~(\ref{eq:master}). Assuming
equal detection and pumping rates for traps 1 and 2, the pumping rates must
satisfy

\begin{itemize}
\item  for two-way pumping with $\chi ^{\leftrightarrow }=\chi ^{\text{in}%
}=\chi ^{\text{out}}:$ 
\begin{equation}
\chi ^{\leftrightarrow }=\frac{\gamma \left\langle n_{1}+n_{2}\right\rangle
+\sigma _{1}\left\langle n_{1}\right\rangle +\sigma _{2}\left\langle
n_{2}\right\rangle }{N_{1}\ +N_{2}\ -\left\langle n_{1}+n_{2}\right\rangle },
\end{equation}

\item  for one-way pumping with $\chi ^{\rightarrow }=\chi ^{\text{in}},$
and $\chi ^{\text{out}}=0:$%
\begin{equation}
\chi ^{\rightarrow }=\frac{\gamma \left\langle n_{1}+n_{2}\right\rangle
+\sigma _{1}\left\langle n_{1}\right\rangle +\sigma _{2}\left\langle
n_{2}\right\rangle }{N_{1}\left( \left\langle n_{1}\right\rangle +1\right)
+N_{2}\left( \left\langle n_{2}\right\rangle +1\right) }.
\end{equation}
\end{itemize}

Note that for $N_{1}\approx N_{2}\gg \left\langle n_{i}\right\rangle ,$ the
two-way pumping rate $\chi ^{\leftrightarrow }$ is larger than the one-way
rate $\chi ^{\rightarrow }$ by a factor of approximately $\left\langle
n_{1}+n_{2}\right\rangle /2.$ In the one-way case, on average one atom is
added for each atom detected or lost to an output coupler, while in the
two-way case, on average {\em all} the trapped atoms are exchanged with the
reservoirs for every loss by detection or output coupling. This difference
has important consequences below.

\section{Quantum state of the traps}

\label{sec:state}

Our main interest in this paper is to find the equilibrium state of the
model just described, under a variety of conditions and explore the
different influences of detection, pumping and collisions. For example, we
naturally expect increasing collisions to reduce the phase coherence and
drive the state towards a narrower number distribution. In preparation,
however, we should first discuss the nature of the entangled state and our
methods for characterizing it more fully.

In earlier studies that consider only detection of the atoms~\cite
{jav96,cir96,nar96,won96,won96a}, the initial state is normally taken as the
product state $\left| \;\right\rangle _{0}=\left| n_{1},n_{2}\right\rangle $
with $n_{1}$ atoms in trap 1 and $n_{2}$ atoms in trap 2. The unnormalized
state following a single detection with phase $\phi $ is found by applying
the field operator $\psi \left( \phi _{1}\right) $: 
\begin{equation}
\quad \left| \;\right\rangle _{1}\propto \left( a_{1}+a_{2}e^{-i\phi
_{1}}\right) \left| n_{1},n_{2}\right\rangle =\sqrt{n_{1}}\left|
n_{1}-1,n_{2}\right\rangle +e^{-i\phi _{1}}\sqrt{n_{2}}\left|
n_{1},n_{2}-1\right\rangle .
\end{equation}
By extension, after $m$ detections the state has the form 
\begin{equation}
\left| \;\right\rangle _{m}=\sum_{k=0}^{m}c_{k}\left( m\right) \left|
n_{1}-m+k,n_{2}-k\right\rangle ,
\end{equation}
where the $c_{k}$ are functions of the phases of all the detected atoms $%
\left\{ \phi _{1},\ldots ,\phi _{m}\right\} .$ If collisions are included,
the state experiences unitary evolution under the Hamiltonian~(\ref{eq:hamil}%
) in between detections and the coefficients $c_{k}\left( m\right) $ are
also functions of time. Here, with the inclusion of pumping the situation is
similar, but as one can continue to detect atoms indefinitely, the entangled
state can become very large\ (note of course that the number of detections $%
m $ can now arbitrarily exceed the initial occupancy of the traps). It
becomes more natural to drop the dependence on $m$ and write the state at
time $t$ as 
\begin{equation}
\left| t\right\rangle \approx \sum_{k=-p}^{p}c_{k}\left( t\right) \left|
n_{1}(t)-k,n_{2}(t)+k\right\rangle .  \label{eq:pumpstate}
\end{equation}
This is an approximate relation because we truncate the sum at some cut-off $%
p.$ This is particularly important numerically as the exact state can become
prohibitively large for calculations. Such a truncation is possible because
the probability of all detections occurring from a single trap is small
(assuming the initial trap occupancies are not wildly different) and hence
the coefficients at the extreme ends of the entangled state are negligible.
In our simulations, we drop terms for which $\left| c_{k}\right| <10^{-12}.$
We characterise the state in terms of the mean number of atoms in each trap,
and the width of the number difference distribution with the natural
definition 
\begin{equation}
\sigma _{n}=\left( \left\langle \left( n_{1}-n_{2}\right) ^{2}\right\rangle
-\left\langle n_{1}-n_{2}\right\rangle ^{2}\right) ^{1/2}.
\end{equation}
Frequently we also wish to describe the phase distribution for which we use
the width~\cite{ban69} 
\begin{eqnarray}
\sigma _{\phi } &=&\left( 1-\left| \left\langle \exp \left( i\phi \right)
\right\rangle \right| ^{2}\right) ^{1/2}  \nonumber \\
&=&\left( 1-\left| \sum_{k}c_{k}^{*}c_{k+1}\right| ^{2}\right) ^{1/2}.
\end{eqnarray}
For a minimum uncertainty state, we have $\sigma _{n}\sigma _{\phi }=1.$
Aside from the evolution of the conditional visibility, our main interest
below is in the behavior of these two measures of the state.

\section{Steady states}

\label{sec:steady}We now turn to finding the steady states of our system.
With thermal pumping schemes however, a genuine steady state is achieved
only for an average over many trajectories. For a given set of parameters,
each trajectory differs not only in the actual relative phase established
between the two traps, but more importantly, in the instantaneous atom
numbers as a function of time. As the trajectory simulation proceeds, the
occupancy of each trap exhibits thermal fluctuations which lead to time
variations in other properties of the system such as the conditional
visibility. Only the time-averaged properties approach a true steady state.
As discussed in section~\ref{sec:ingreds}, we treat two cases: two-way
pumping in which atoms may be exchanged between the bath and trap in both
directions, and one-way pumping in which atoms can only move from the bath
into the trap. We begin with representative plots of the visibility as a
function of time for a single trajectory with no collisions $\left( \kappa
=0\right) .$ The visbility for two-way pumping is shown in Fig.~\ref
{fig:twoway}a. There are initially $n=100$ atoms in each trap and the
pumping rate is chosen to balance the detection rate. On average, $2n$ atoms
are detected in a time $\gamma t=1.$ The visibility is extremely noisy with
frequent fluctuations of order 1. Our simulations show the occupancies of
the traps also display large fluctuations as would be expected for coupling
to thermal baths In particular, a zero in the visibility is always
associated with a zero in one or other of the atom numbers. The visibility
for a typical trajectory with one-way pumping is shown in Fig.~\ref
{fig:twoway}b. In this case, there are again large fluctuations but on a
much longer time scale. This difference has a simple origin mentioned
earlier: for the one-way case, on average one atom is added to the system
for each atom detected and so all the atoms are replaced once in time $%
\gamma t=1$. In the two-way case, $n$ atoms are exchanged with the baths for
each atom detected, and so in $\gamma t=1$, all the atoms are replaced $n$
times over and we expect a correspondingly shorter time scale for the
fluctuations. As a comparison, in Fig.~\ref{fig:twoway}c we show a
trajectory for a ``regular'' pumping model in which atoms are dripped into
the trap at a constant rate to replace those lost by detection. In this
case, the collisional rate is $\kappa =0.5\gamma ,$ but the visibility shows
a much improved response than in the (collisionless) thermal pumping cases,
indicating the severe influence of the thermal pumping.

In fact, the visibility is degraded by the number fluctuations in two
distinct ways. When the occupancy of one of the traps falls due to a
fluctuation to within a few times $\sigma _{n}$ of zero, the extreme terms
in the entangled state are removed, the number distribution narrows and the
visibility falls. In particular, if one of the traps is completely emptied
(as occurs several times in Figs.~\ref{fig:twoway}a and b,) the state is
then a pure number difference state and any relative phase is completely
destroyed. The visibility can of course be restored once fluctuations
increase the atom number again, but there is no relation between the new
relative phase and the phase before the trap was emptied.

Even when both traps have $\left\langle n\right\rangle \gg \sigma _{n}$, the
visibility is reduced according to the number difference between the traps.
This is obvious---if one trap has significantly more atoms than the other,
then we can predict with better than 50~\% accuracy from which trap the next
atom will be detected, and the visibility must fall accordingly. We can
calculate this effect simply as follows. In our system, the atom numbers
experience thermal fluctuations in {\em time} due to the pumping. Suppose
for a moment we have a different situation in which there is no pumping, and
we perform a series of detection runs with a thermal distribution in the 
{\em initial} trap numbers and measure the visibility after a well-defined
relative phase has been set up (but before the traps are significantly
depleted). If we picture the condensates as coherent states with some
relative phase: 
\begin{equation}
\left| \;\right\rangle _{1}=\sqrt{n_{1}}\exp \left( i\phi _{1}\right)
,\qquad \left| \;\right\rangle _{2}=\sqrt{n_{2}}\exp \left( i\phi
_{2}\right) ,
\end{equation}
the expected visibility is just $\beta =2\sqrt{n_{1}n_{2}/}\left(
n_{1}+n_{2}\right) $ which is a familiar expression for optical fields.
Defining the relative occupancy 
\begin{equation}
f=\frac{n_{1}-n_{2}}{n_{1}+n_{2}},  \label{eq:fdefn}
\end{equation}
we have 
\begin{equation}
\beta \left( f\right) =\sqrt{1-f^{2}}.  \label{eq:betafrel}
\end{equation}
More correctly, we should derive Eq.~(\ref{eq:betafrel}) directly from the
entangled state description of the twin-trap system. In general, the fringe
visibility is given by 
\begin{equation}
\beta =\left| g^{\left( 1\right) }\right| \sqrt{1-f^{2}},
\end{equation}
where $\left| g^{\left( 1\right) }\right| $ is the normalized correlation
function~\cite{wal94}. While in general, $\left| g^{\left( 1\right) }\right| 
$ is not easily evaluated for the entangled states with which we are
concerned, it can be shown for example that for the projected two-mode
coherent state~\cite{mol96} $\left| \alpha ,\beta \right\rangle
=\sum_{k=0}^{N}\alpha ^{k}\beta ^{N-k}/\sqrt{k!\left( N-k\right) !}\left|
k,N-k\right\rangle ,$ which is the most natural expression of a state with
relative phase with fixed total atom number $N,$ $\left| g^{\left( 1\right)
}\right| $ tends to unity in the limit of large $N,$

Figures~\ref{fig:scatter}a and~b test Eq.~(\ref{eq:betafrel}) in the form of
scatter plots of the points $\left( f\left( t\right) ,\beta \left( t\right)
\right) $ for two-way and one-way pumping respectively with the same
parameters as Figs~\ref{fig:twoway}a and b. The prediction~(\ref{eq:betafrel}%
) is indicated by the black squares in each plot. The correlation is clearly
much stronger in the one-way case. This difference is entirely due to the
difference in time scales discussed above. If we are to have a well-defined
phase, the number must be partially uncertain. Indeed, we see later that in
the presence of collisions the average state has moderate number-squeezing
but with a variance of the same order as a coherent state. So for a
reasonable visibility we should require an entangled state of order $\sigma
=O\left( 2\sqrt{n}\right) $\ terms. Such an entangled state is set up by the
same number of detections and requires a time of order $\tau _{\text{e}}=1/%
\sqrt{n}\gamma .$ Now for two-way pumping, the time scale for replacement of
all the atoms once over is $\tau _{\text{r}}=1/n\gamma \ll \tau _{\text{e}}.$
Hence, the exchange of atoms with the baths occurs faster than an entangled
state of a particular phase can be constructed and we may expect a reduced
visibility. There can be only a weak correlation between the instantaneous
visibility and the instantaneous relative occupancy $f$ , and the visibility
is generally lower than the optimum given by Eq.~(\ref{eq:betafrel}). With
one-way pumping however, the time scale for replacement of all the atoms is
larger by a factor $n.$ The visibility is able to keep up with the drift in
number and is then limited only by~Eq.~(\ref{eq:betafrel}).

We can also calculate the mean visibility over time $\overline{\beta },$ for
an arbitrary pair of mean atom numbers $\left\langle n_{1}\right\rangle $
and $\left\langle n_{2}\right\rangle .$ Again, we think of an ensemble of
runs with no pumping and a thermal distribution of initial states $\left|
n_{1}\right\rangle $ and $\left| n_{2}\right\rangle .$ The mean visibility
over many runs is the weighted average of $\beta \left( f\right) $ over the
probability distribution [see Eq.~(\ref{eq:fdefn})] 
\begin{equation}
P_{f}\left( f\right) =\int \delta \left( f-\frac{n_{1}-n_{2}}{n_{1}+n_{2}}%
\right) P_{\bar{n}_{1}}\left( n_{1}\right) P_{\bar{n}_{2}}\left(
n_{2}\right) \;dn_{1}dn_{2},
\end{equation}
where $P_{\bar{n}_{i}}\left( n_{i}\right) =-\log \left( \gamma _{i}\right)
\gamma _{i}^{n_{i}}$ are the probability distributions (in the continuous
limit) of atom number for thermal distributions with mean number $\bar{n}%
_{i} $ and $\gamma _{i}=\bar{n}_{i}/\left( \bar{n}_{i}+1\right) .$ For the
case where the mean numbers are the same, $P\left( f\right) $ is uniform and 
$\bar{\beta}=\pi /4~$\cite{gra97}$.$ Otherwise we find 
\begin{equation}
\bar{\beta}=\frac{2\pi \log \left( \gamma _{1}\right) \log \left( \gamma
_{2}\right) }{\left[ \log \left( \gamma _{1}/\gamma _{2}\right) \right] ^{2}}%
\left( \frac{1}{\sqrt{1-\left( \frac{\log \left( \gamma _{1}/\gamma
_{2}\right) }{\log \left( \gamma _{1}\gamma _{2}\right) }\right) ^{2}}}%
-1\right) .
\end{equation}
which for $\bar{n}_{1},$ $\bar{n}_{2}\gg 1,$ gives 
\begin{equation}
\bar{\beta}\simeq \frac{\pi \sqrt{p}}{\left( 1+\sqrt{p}\right) ^{2}},
\label{eq:betamean}
\end{equation}
with $p=\bar{n}_{1}/\bar{n}_{2}.$ In the pumped twin-trap setting, we also
have thermal distributions in the atom number which occur not from run to
run, but over time in a single trajectory, so it is reasonable to hope that
the above analysis may still apply.

In Fig.~\ref{fig:pdepend}, we show the average visibility for a thermal
distribution as a function of the mean atom number ratio $p$ given by Eq.~(%
\ref{eq:betamean}). The plotted points show the time-averaged visibility
calculated from trajectory simulations with one-way pumping and $\bar{n}_{1}+%
\bar{n}_{2}=200.$ Error bars are shown at 1 standard deviation. As expected,
the mean visibility falls with increasing disparity in the mean atom number.

\subsection{One-way pumping and output couplers}

We have seen that the one-way pumping process shows significantly higher
visibility than the two-way pumping. From this point on, we restrict our
attention to the one-way model and add the effect of an output coupler from
each trap. In our results we find two distinct regimes according to the
length of the simulations. Figure~\ref{fig:longbeta} shows the visibility as
a function of time averaged over 200 trajectories for a simulation with $%
\kappa =0,$ $\nu _{i}=0$ and an initial state $\left|
n_{1},n_{2}\right\rangle =\left| 100,100\right\rangle .$ The one-way pumping
rate was chosen to maintain the mean population at $n=100$ in each trap.
There are clearly two regimes: for $\gamma t\precapprox n,$ the mean
visibility shows a steady decline, while for $\gamma t\succapprox n,$ the
visibility tends to a steady-state value of $\pi /4$ consistent with the
calculation of the previous section. In the initial stage of a particular
run, the populations of the traps become decorrelated due to the thermal
nature of the pumping until they are completely uncorrelated and the
time-averaged visibility is $\pi /4.$ The time for this decorrelation varies
from run to run, having a characteristic length of $\gamma t\approx n.$ Thus
the average over many trajectories shows a gradual decline until all members
of the ensemble are likely to be decorrelated. We are thus led to examine
the behavior of the system in the two regimes $\gamma t\ll n,$ when the trap
populations are likely to be quite close, and $\gamma t\gg n,$ when there is
no correlation between the populations. We treat these two cases in turn. In
all cases, we start our simulations with the initial state $\left|
n_{1},n_{2}\right\rangle =\left| 100,100\right\rangle $ and calculate
quantities averaged over 200 trajectories.

\begin{itemize}
\item  $\gamma t\ll n:$ For the short-time regime, we arbitrarily choose $%
\gamma t=4$ to show results. Figure~\ref{fig:paramshort}a shows the
visibility as a function of $\kappa $ again averaged over five trajectories.
As expected, the visibility decreases with increasing collisions which
increasingly disrupt the relative phase~\cite{won96}. We have also performed
simulations with a range of output couplings from $\nu _{i}=0$ to $\nu _{i}=$
$\gamma $. This leads to a small decrease in $\beta $ (of less than 0.025
for the strongest coupling). This effect is simply a result of the fact that
the pumping rate is increased to balance the additional loss of atoms and so
the trap populations decorrelate faster. The nature of the average state of
the system is indicated in Fig.~\ref{fig:paramshort}b. Shown are the widths
of the number distribution $\sigma _{n}$ (dotted line) and phase
distribution $\sigma _{\phi }$ (dot-dashed) and $\rho $: the root mean
square of the product of the two (solid). The filled circles denote the
actual simulations performed. The number distribution clearly narrows
strongly with increasing collisional rate while the phase distribution
spreads as collisions degrade the relative phase. The simulations with
non-zero output coupling (not shown in Fig.~\ref{fig:paramshort}b) produced
a reduction of less than 5 \% in the number width and no discernible change
in the other parameters. Note that for zero collisions, the product of the
widths (solid) is unity indicating a minimum uncertainty state. Further, for
all values of $\kappa ,$ the number width $\sigma _{n}<\sqrt{\bar{n}}=10,$
which is the width we would expect if the state was a projection of a
coherent state onto a basis of fixed total atom number. The real state is
thus quite strongly number squeezed. This in consistent with recent analytic
work by Dunningham {\em et al~}\cite{dun97,dun97a} using a Bose-broken
symmetry model. They show that in the limit of a large collisional rate, the
true state of the condensate is the amplitude-squeezed state that minimizes
number fluctuations.

\item  $\gamma t\gg n:$ In the large time regime, the mean visibility has no
dependence on the output coupling rate---once the atom numbers are
completely uncorrelated, the precise rate at which atoms enter the trap is
irrelevant. The number and phase widths shown in Figure~\ref{fig:paramlong}
show very similar trends to the short time case. Note that even with the
uncorrelated trap numbers, the state is still minimum uncertainty for $%
\kappa =0,$ indicating that the pumping rate is low enough for the
visibility to adjust to changes in number. Again, other simulations showed
that the only effect of output coupling was to reduce the number width by a
few percent.
\end{itemize}

\section{Application to collapses and revivals}

\label{sec:collapses}

In this final section, we consider the application of pumping processes to
the interesting phenomenon of collapses and revivals in the relative phase.
Several authors have shown that if a relative phase is prepared by detection
and the entangled state subsequently evolves purely under the influence of
the interatomic collisions, the visibility of the phase experiences
recurrent collapses and revivals of period $\pi /\kappa $ due to the
differential rate of phase rotation in the entangled state~\cite
{lew96,won96a,wri96,wri97}. A demonstration of collapses and revivals of the
phase, perhaps through light scattering experiments, would be a significant
result in BEC\ physics. It is interesting to consider how the collapses and
revivals are affected by pumping and leaking of atoms through the traps.
Naively, we might expect that the oscillations would be destroyed by the
time all the atoms had been replaced a few times over. In fact, we have
found the collapses and revivals to be remarkably robust to pumping
processes.

In Fig.~\ref{fig:collapse}a, we show the visibility for a single trajectory 
{\em without} pumping or output coupling in which there are initially 200
detections from a total of 1000 atoms, followed by a period in which the
system evolves only under the influence of collisions with $\kappa =0.25$.
The oscillations in the visibility are clear. Fig.~\ref{fig:collapse}b shows
a trajectory with the same number of detections and collision strength, but
with a continual flushing of the trapped atoms by pumping and output
coupling. On average, all the atoms are replaced in a period $\gamma t=1$
and the atom numbers exhibit large fluctuations (Fig.$~$\ref{fig:collapse}%
c). Despite this, the collapse and revivals persist for a considerable
period and only disappear when the atom number in trap 2 (thick line in Fig.$%
~$\ref{fig:collapse}c) approaches zero at $\gamma t\approx 250.$ If the
trajectory is such that neither atom number approaches zero, the
oscillations may continue much longer still. As the simulation progresses
however, while the period of the revivals is unchanged, the peaks
broaden---the collapse time increases. This is associated with the gradual
reduction in the width of the number difference of the initial entangled
state due to the repeated addition and removal of atoms indicated in Fig.$~$%
\ref{fig:collapse}d. Essentially, every addition or removal of an atom
through an output coupler tends to drive the state to a narrower number
distribution. In their treatment of collapses and revivals for a {\em fixed}
number of atoms, Wong {\em et al}.$~$\cite{won96a} have shown that the
visibility during a collapse should decay according to 
\begin{equation}
V\propto \exp \left( -2\sigma _{A}^{2}\kappa ^{2}t^{2}\right) ,
\label{eq:decay}
\end{equation}
where $\sigma _{A}$ is the width of a Gaussian approximation to the
coefficients 
\begin{equation}
{\cal A}\left( k\right) =\left| c_{k}c_{k-1}\right| \sqrt{\left(
n-k+1\right) \left( n-m+k\right) },
\end{equation}
and there have been $m$ detections from an initial state of $n$ atoms in
each trap. For a broad distribution, and $m\ll n,$ to lowest order we have $%
{\cal A}\left( k\right) \propto \left| c_{k}^{2}\right| ,$ so that in our
notation $\sigma _{A}\approx \sigma _{n}/2.$ The black squares in Fig.$~$\ref
{fig:collapse}d are estimates of $\sigma _{n}$ calculated from the collapse
widths in Fig.$~$\ref{fig:collapse}b using Eq.~(\ref{eq:decay}). The
agreement with the directly measured values for the width of the number
distribution (solid line) confirms that the increase in the collapse time is
due purely to the change in $\sigma _{n.}$

Figure~\ref{fig:collapse}b also shows a variation in the height of the
visibility peaks. Note that the variation is not monotonic, an effect we
have found to be generally true. A natural guess is that the peak heights
are associated with the relative occupancy $f=\left( n_{1}-n_{2}\right)
/\left( n_{1}+n_{2}\right) ,$ which we found led to a maximum visibility for
systems where the detections are {\em not} stopped in section~\ref
{sec:steady}. We have tested this using scatter graphs of the peak
visibility similar to those in Fig.~\ref{fig:scatter}. We find a moderate
confirmation of the connection. In cases for which the {\em minima} of the
visibility remain small, there is a strong correlation between the peak
visibility and the quantity $f$. In other cases, such as that in Fig.$~$\ref
{fig:collapse}b for $\gamma t>150,$ for which the minima are significantly
greater than zero, the correlation is poor and we conclude that the pumping
process has produced an additional degradation of the state beyond that
implied simply by the mean number difference.

\section{Conclusion}

In this paper, we have studied the steady-state behavior for two pumping
scenarios to show how an ongoing measurement process can generate a phase
coherence between atoms derived from thermal baths, even in the presence of
phase diffusion due to atomic collisions within the traps. We find important
qualitative differences between systems with two-way and one-way pumping,
the phase coherence being substantially improved for the one-way case.
Systems displaying collapses and revivals of the condensate phase should
provide an opportunity for examining the time-dependent effects induced by
pumping. We remark finally that a natural extension to our model would be
the inclusion of extra trap levels that would allow for line narrowing and
genuine laser action.

\acknowledgments

The authors thank Tony Wong and Matthew Collett for useful discussions. We
acknowledge support of the Marsden fund of the Royal Society of New Zealand,
the University of Auckland Research Committee and the New Zealand Lotteries'
Grants Board.

\begin{figure}[tbp]
\caption{Geometry of pumped twin-trap system. The straight solid arrows
indicate the detection at rate $\gamma$; the dotted arrows, the exchange of
atoms with the reservoirs; the curved arrows, output coupling of the trapped
atoms.}
\label{fig:geom}
\end{figure}

\begin{figure}[tbp]
\caption{a)-b) Visibility $\beta$ as a function of time for a thermally
pumped system with mean occupancy $n=100$ for each trap and $\kappa=0$. a)
Two-way pumping, b) one-way pumping. c) Visibility for a regularly pumped
model with $\kappa=1.0$. }
\label{fig:twoway}
\end{figure}

\begin{figure}[tbp]
\caption{Scatter plot of $\beta$ as a function of relative occupancy $f$ for
a) two-way, and b) one-way pumping. There are 10000 points shown. Black
squares indicate the relation Eq.~(\ref{eq:betafrel})}
\label{fig:scatter}
\end{figure}

\begin{figure}[tbp]
\caption{Mean visibility $\bar{\beta}$ as a function of the atom number
ratio $p$. Error bars indicate 1 standard deviation errors in time-averaged
simulations.}
\label{fig:pdepend}
\end{figure}

\begin{figure}[tbp]
\caption{Visibility for one-way thermal pumping with no collisions and no
output coupling. The mean atom number in each trap is 100.}
\label{fig:longbeta}
\end{figure}

\begin{figure}[tbp]
\caption{Averaged state parameters as a function of collision rate for
one-way pumping in short-time regime. a) Visibility and b) $\sigma_n $
(dashed), $\sigma_\phi$ (dot-dash), and $\rho$ (solid).}
\label{fig:paramshort}
\end{figure}

\begin{figure}[tbp]
\caption{Averaged state parameters as a function of collision rate for
one-way pumping in long-time regime: $\sigma_n $ (dashed), $\sigma_\phi$
(dot-dash), and $\rho$ (solid).}
\label{fig:paramlong}
\end{figure}

\begin{figure}[tbp]
\caption{a) Visibility for collapses and revivals of relative phase with no
pumping or output. Initially 200 detections were made from 1000 atoms. b)
Visibility, c) atom numbers and d) $\sigma_n$ for the same parameters with
pumping and output rates such that all atoms are replaced on average once in
a period $\gamma t=1$.}
\label{fig:collapse}
\end{figure}

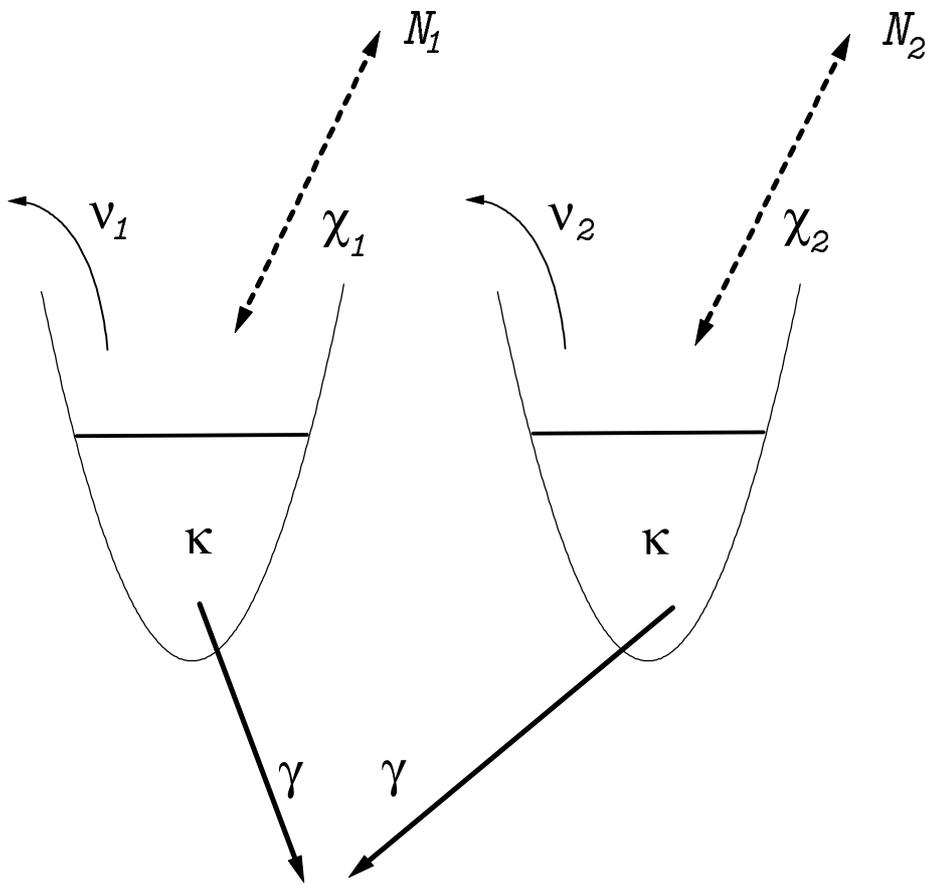
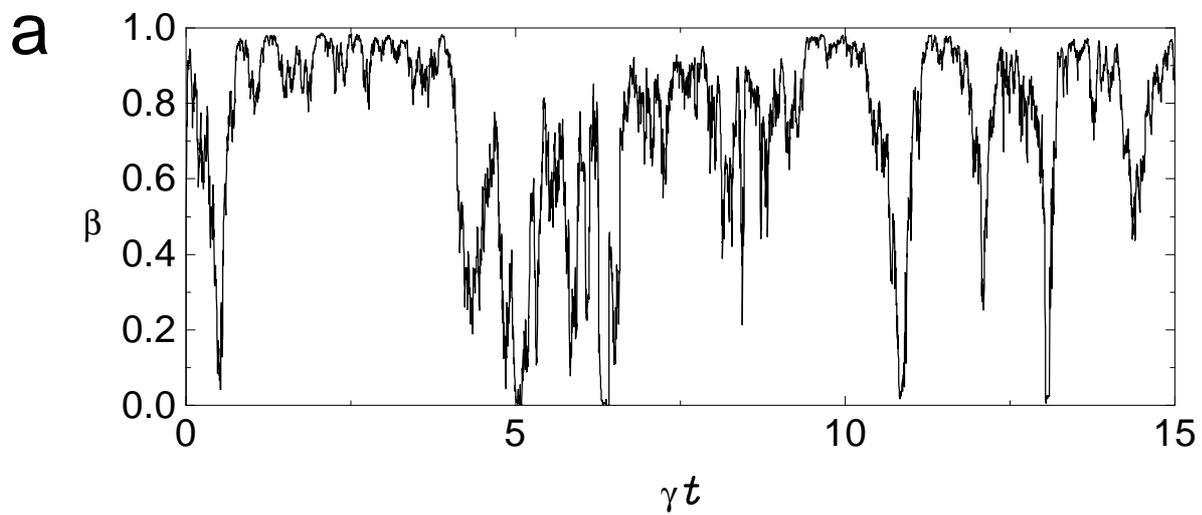
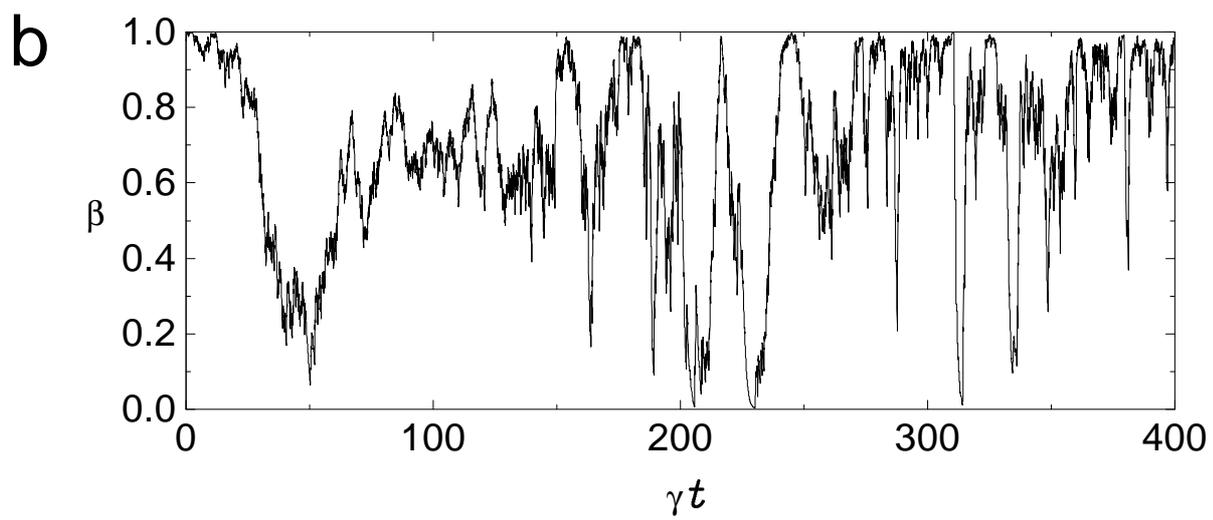
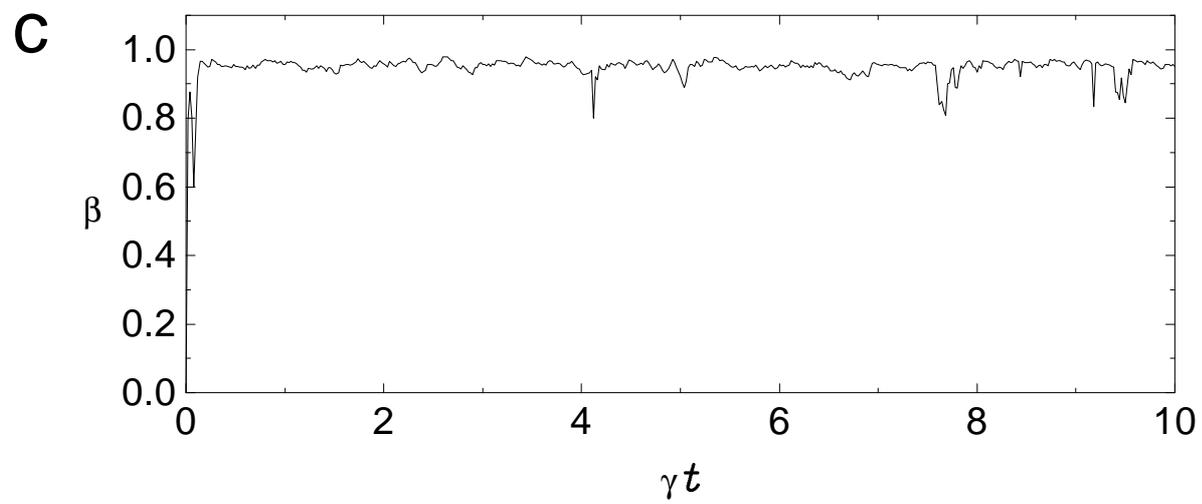
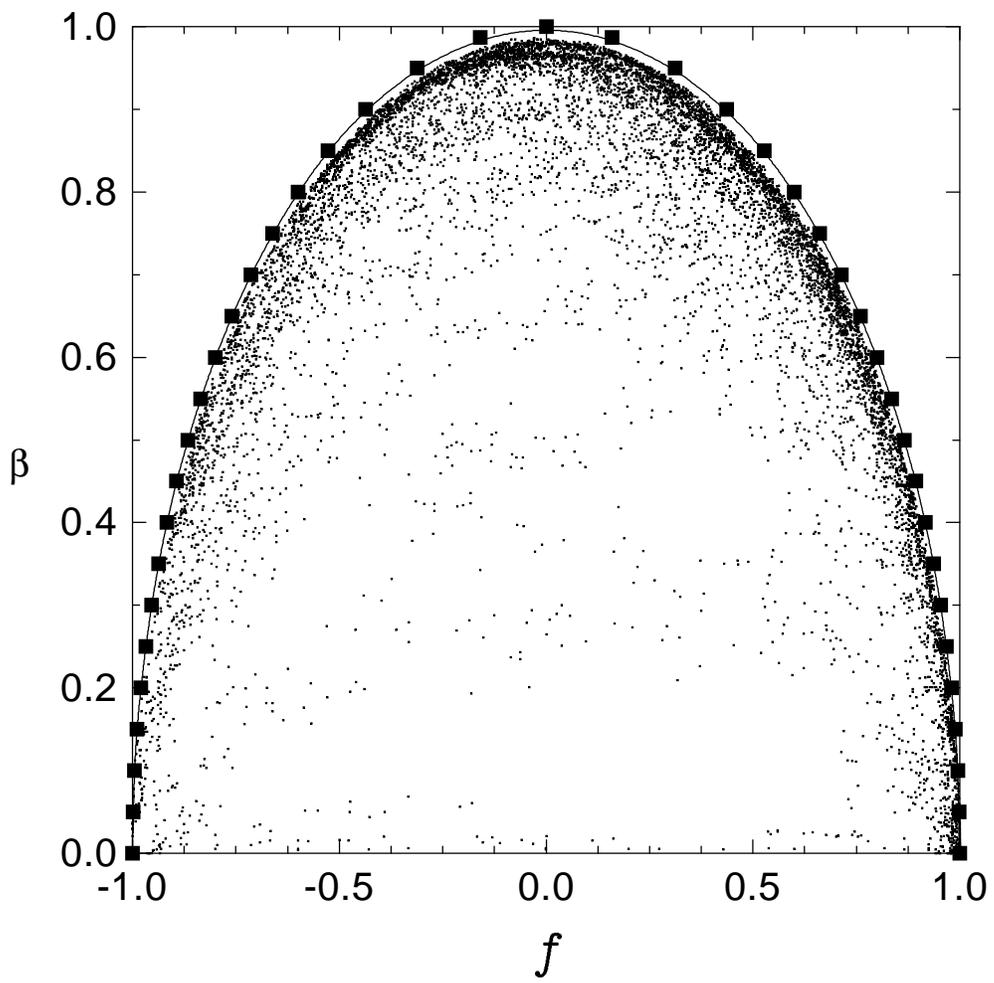
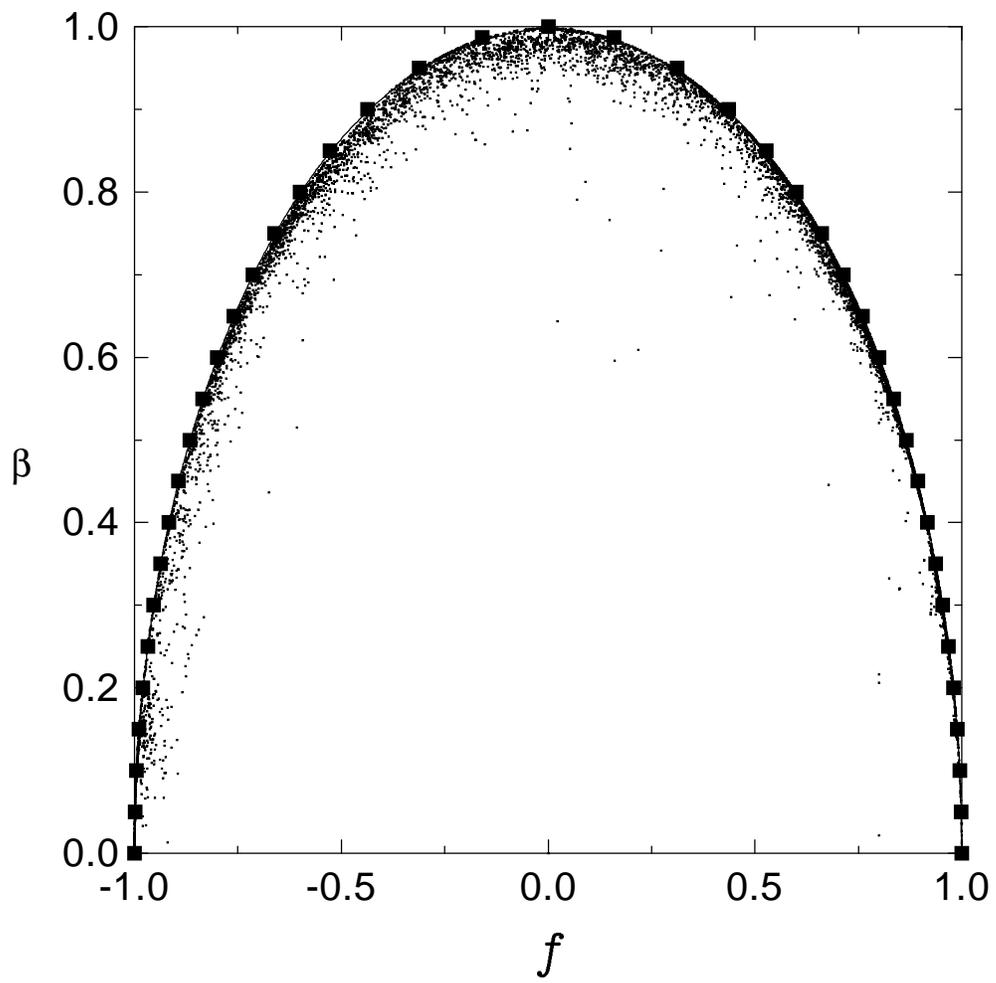
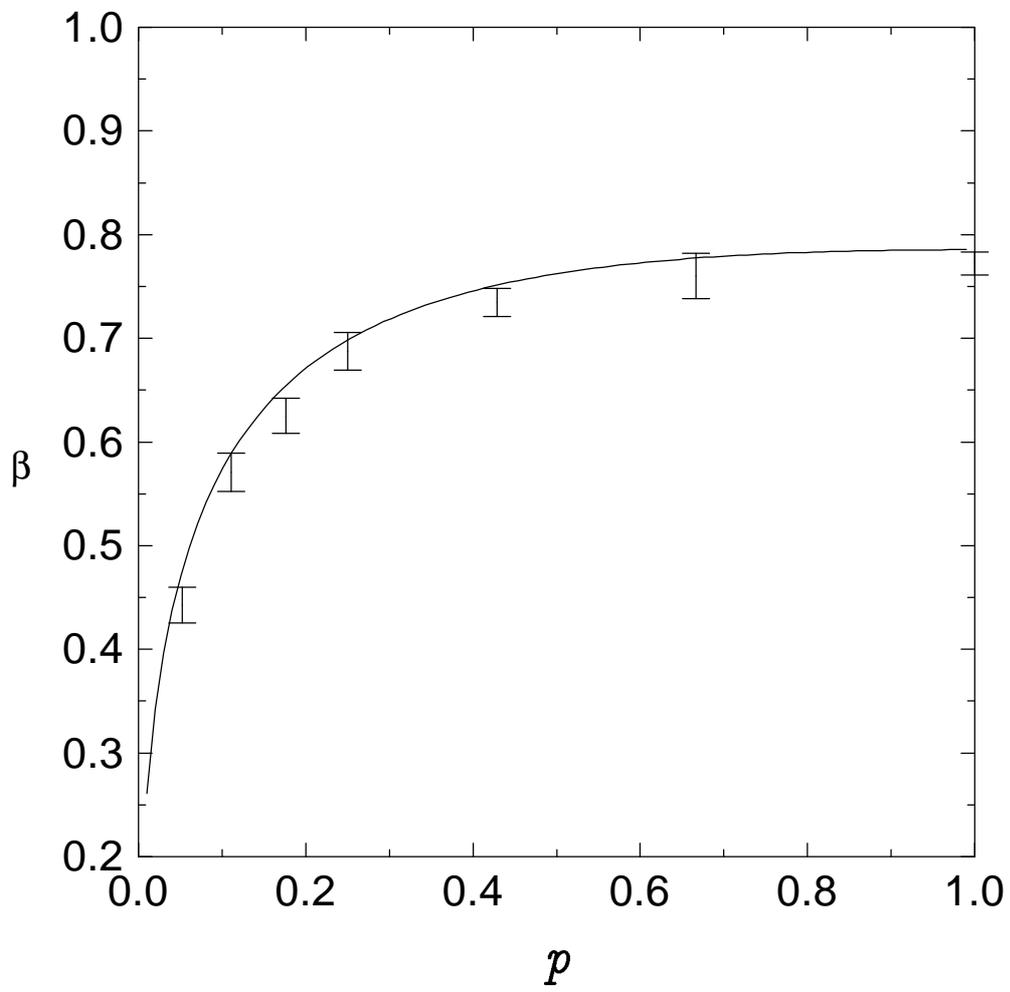
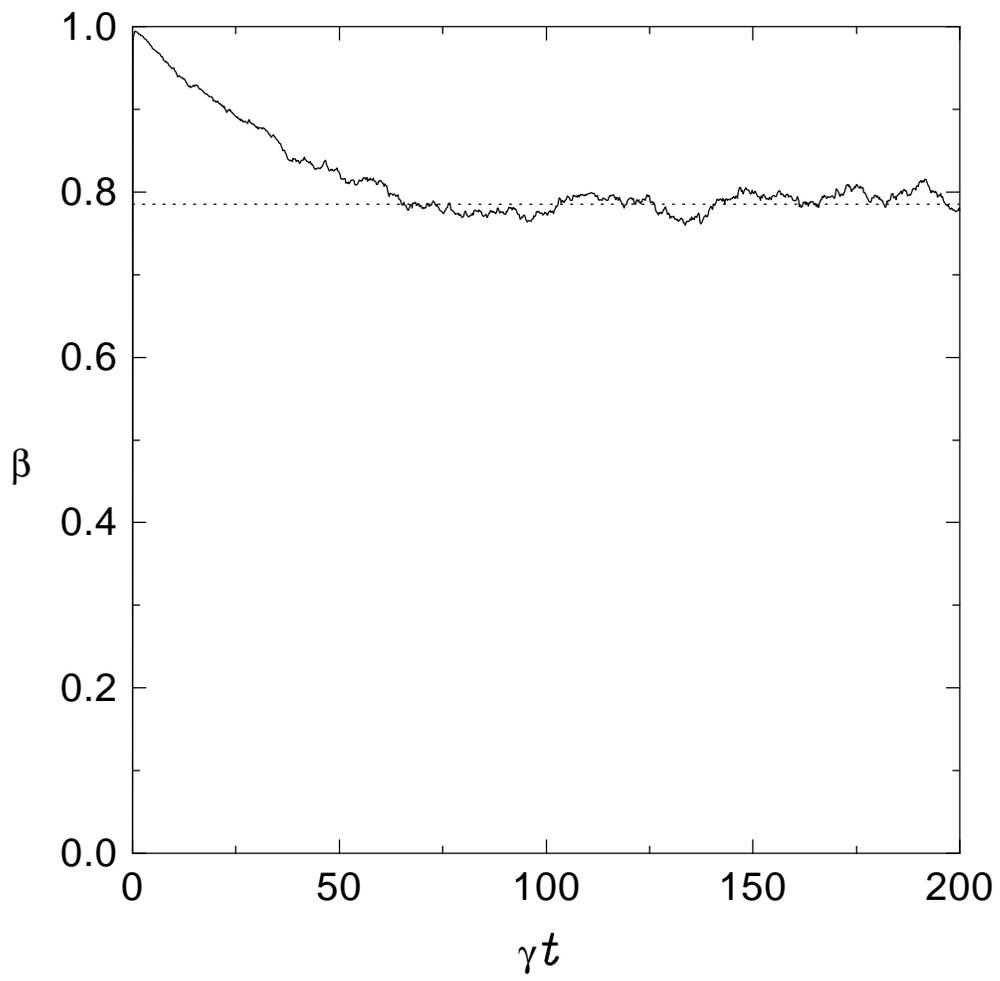
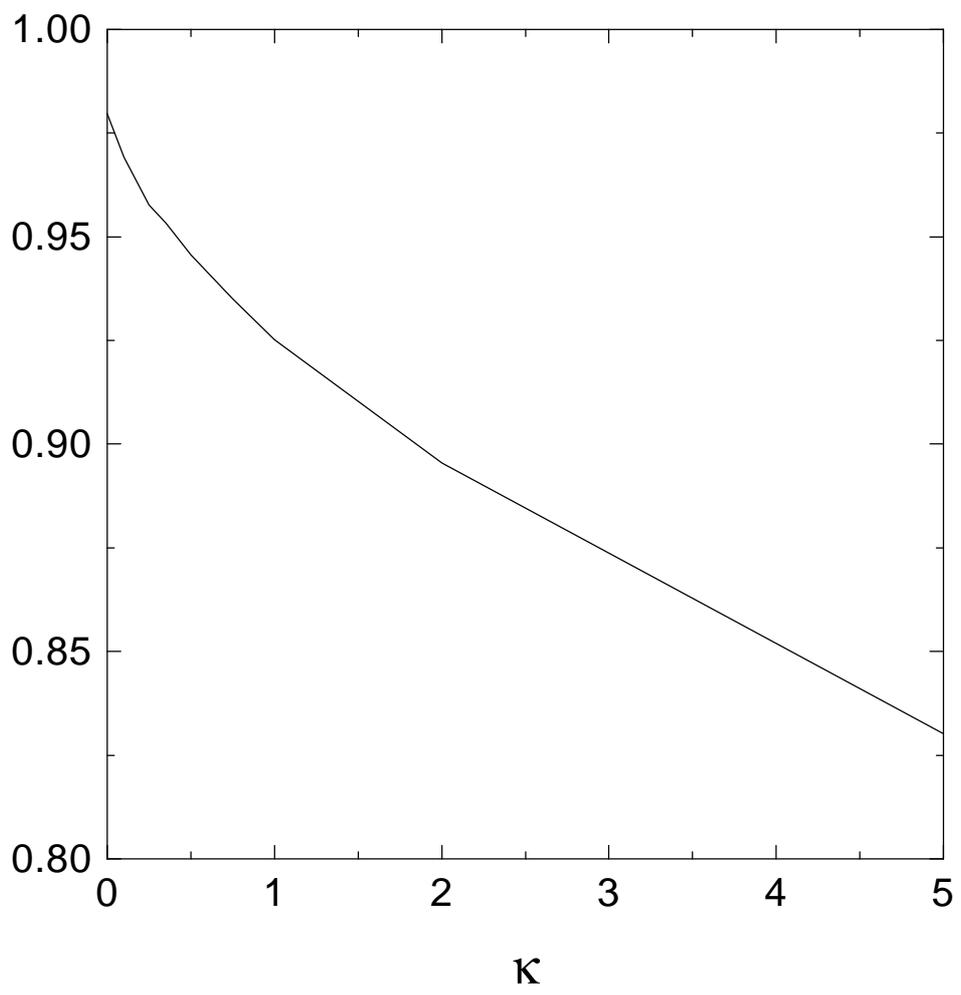
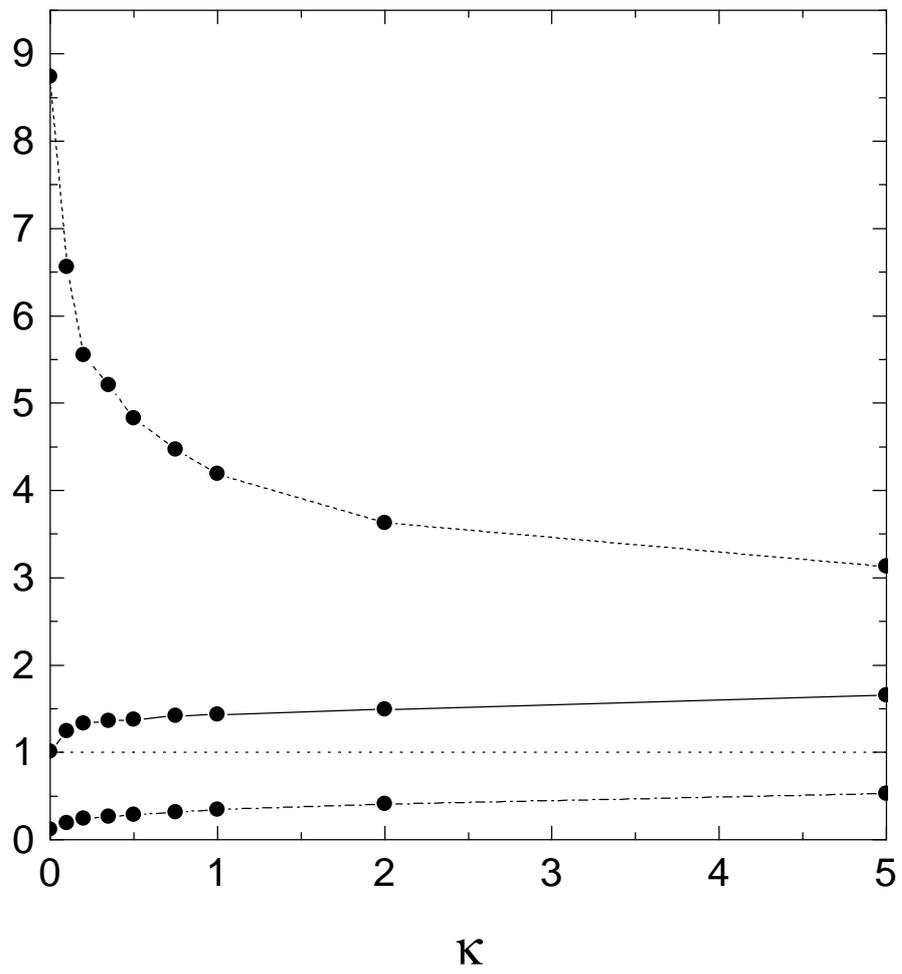
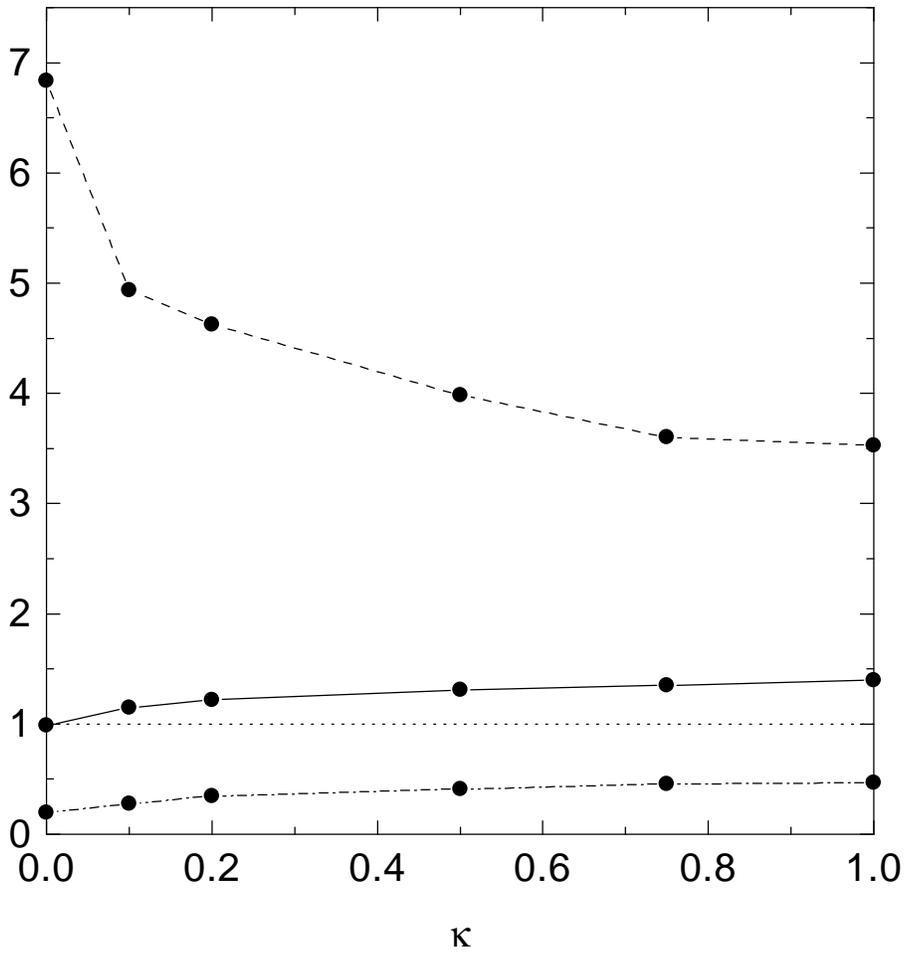
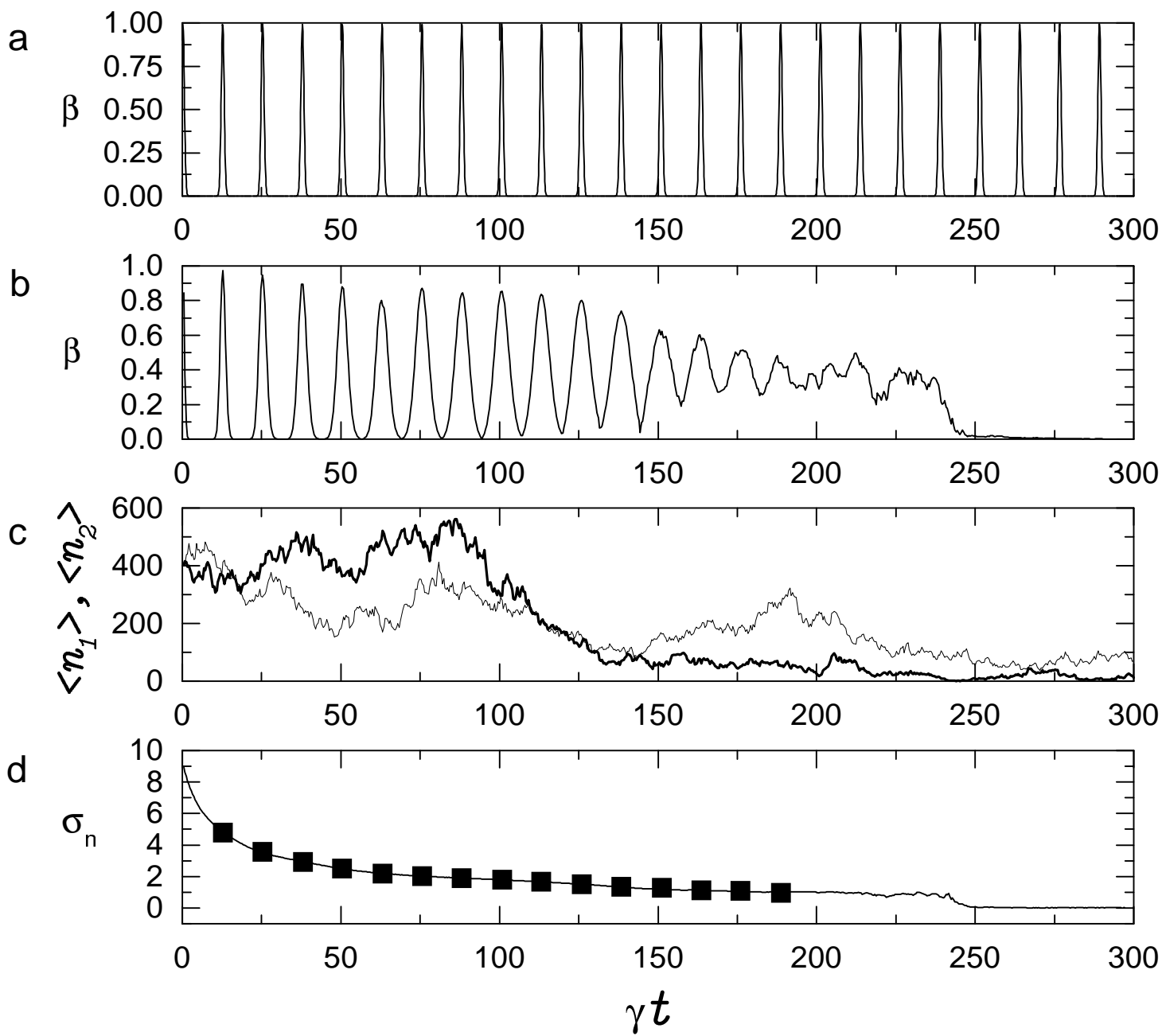
\end{document}